\documentclass[twocolumn,prb,showpacs,floatfix]{revtex4}
\usepackage{graphicx}
\usepackage{bm}
\usepackage[dvipdfm]{hyperref}
\newcommand{\be}{\begin{equation}}
\newcommand{\ee}{\end{equation}}
\newcommand{\bea}{\begin{eqnarray}}
\newcommand{\eea}{\end{eqnarray}}

\newcommand{\half}{\textstyle\frac{1}{2}}

\begin{document}

\title{\textbf{Magnon and Hole Excitations in the
Two-Dimensional Half-filled Hubbard Model}}

\author{
        Weihong Zheng$^{1}$,
        Rajiv R.~P.~Singh$^{2}$,
        Jaan Oitmaa$^{1}$,
        Oleg P. Sushkov$^{1}$
        Chris J.~Hamer$^{1}$
}

\affiliation{ $^1$School of Physics, University of New South Wales, Sydney, NSW 2052, Australia\\
$^2$Department of Physics, University of California, Davis, CA 95616\\
}

\date{\today}

% 75.10.Jm Quantized spin models

%%%%%%%%%%%%%%%%%%%%%%%%%%%%%%%%%%%%%%%%%%%%%%%%%%%%%%%%%%%%%%%%%%%%
\begin{abstract}
Spin and hole excitation spectra and spectral weights
are calculated for the half-filled Hubbard model, as a function of $t/U$.
We find that the high energy spin spectra are sensitive to charge fluctuations.
The energy difference $\Delta(\pi,0)- \Delta(\pi/2,\pi/2)$,
which is negative for the Heisenberg model, changes
sign at a fairly small $t/U\approx 0.053(5)$.
The hole bandwidth is
proportional to $J$, and considerably larger than in the
$t$-$J$ models. It has a minimum at ($\pi/2,\pi/2$) and a very weak
dispersion along the antiferromagnetic zone boundary.
% The measured spin-spectra in La$_2$CuO$_4$
% lead to the estimates $U=3.1{\rm eV}$, $t=0.35{\rm eV}$.
A good fit to the measured spin spectra in  La$_2$CuO$_4$ at $T=10K$ is obtained with
the parameter values $U=3.1{\rm eV}$, $t=0.35{\rm eV}$.
%at $10K$ can be well fitted by the Hubbard model with $U/t=8.75\pm 0.7$.
% but imply that the effective $U/t$ ratio increases from $8.75\pm 0.7$
%at $T=10K$ to $10.5\pm 0.7$ at 295K.
\end{abstract}
%%%%%%%%%%%%%%%%%%%%%%%%%%%%%%%%%%%%%%%%%%%%%%%%%%%%%%%%%%%%%%%%%%%%%%%%%

\maketitle

%%%%%%%%%%%%%%%%%%%%%%%%%%%%%%%%%%%%%%%%%%%%%%%%%%%%%%%%%%%%%%%%%%%%%%%%%
\section{Introduction}
%%%%%%%%%%%%%%%%%%%%%%%%%%%%%%%%%%%%%%%%%%%%%%%%%%%%%%%%%%%%%%%%%%%%%%%%%
Underdoped phases of high temperature superconducting materials
and the nature of the metal insulator transition upon doping a
Mott-insulating antiferromagnet remain central
topics of research in condensed matter physics. Some puzzles
extend all the way to the undoped stoichiometric insulating materials.
Results such as the
antiferromagnetic zone-boundary magnon excitations probed in inelastic neutron
scattering \cite{Radu,Kampf},
the two-magnon excitations probed in Raman scattering \cite{raman}
and the one-hole excitations probed in angle-resolved photoemission
spectroscopy \cite{ARPES}
contiune to surprise us. The question of whether some of these
anomalies are connected to the pseudogap phase of the weakly doped
materials remains a topic of debate.

An important question is the extent to which
conventional approaches, based on  ordered
antiferromagnetic phases, can explain the observed spectra and
 spectral weights and to what extent the interpretation of data
necessitates the introduction of novel ideas such as
spin-liquids and spin-charge separation. The low energy long-wavelength
spin excitations of the antiferromagnet are well described by the
non-linear sigma model \cite{CHN}.
However, the high energy zone-boundary spin excitations
necessarily require a microscopic lattice model. The
antiferromagnetic insulator, without charge fluctuations,
is represented by the Heisenberg model, and the excitation
spectrum of this model has been the subject of several controlled numerical
studies \cite{sandvik,gelfand,HAF}.
It is clear that the antiferromagnetic zone-boundary
spectrum of La$_2$CuO$_4$ does not agree with that of the Heisenberg
model. In particular, in the Heisenberg model, the magnon
energy difference $\Delta(\pi,0)-\Delta(\pi/2,\pi/2)$ is negative
but it is found to be positive for La$_2$CuO$_4$. This result will
be worse if second neighbor antiferromagnetic interactions are
included. It has been suggested that one way to reconcile the
difference is by invoking ring-exchange terms \cite{Radu,Kampf,troyer,ring},
which arise due to charge fluctuations \cite{girvin}.

Here, we present systematic numerical calculations of
the magnon and hole spectra and spectral weights of the Hubbard model as
a function of $t/U$. First, we focus on the magnons.
Earlier the magnon spectra were studied by
mean-field theory \cite{MFT} and by a Quantum Monte Carlo
Simulation combined with the Single Mode Approximation \cite{SMA}, neither
of which are expected to be quantitatively accurate for small $t/U$.
Our calculations show that the zone-boundary magnon energies
are very sensitive to charge fluctuations and
the difference $\Delta(\pi,0)-\Delta(\pi/2,\pi/2)$ changes sign at a relatively
small $t/U$ value of $0.053 (5)$. The magnon spectra of La$_2$CuO$_4$
and the spectral weights are
well described by the Hubbard model as discussed below.
%We also compare our results
%with the neutron scattering spectra on two other
%square-lattice antiferromagnetic materials with an
%exchange constant an order of magnitude smaller than La$2$CuO$_4$. These are
%(CuDCOO)$_2\cdot 4$D$_2$O (CFTD) \cite{Ronnow}, and
%SrCuOCl \cite{Kim}. The latter belongs to the Cuprate family and has
%a weakly coupled spin system due to its unusual geometry.

The calculated hole spectra, on the other hand, are qualitatively similar to
previous theoretical studies of Hubbard and t-J models \cite{dagotto,bulut}.
The hole-bandwidth is suppressed at large $U$ by a factor of $t/U$,
although we find it to still be much larger than in the corresponding $t$-$J$ models.
The minimum is at ($\pi/2,\pi/2$) with a weak dispersion along the
antiferromagnetic zone boundary. Hence, these results cannot be used to
fit the observed ARPES spectra in the undoped cuprate materials \cite{ARPES}.
Although same-sublattice hopping terms can allow better fits to
the dispersion, the anomalous spectral weights remain harder to
explain\cite{ARPES}.
%and point to the importance of even small same-sublattice hopping terms
%in the Hamiltonian for the holes, which are evidently not as important for the spin
%dispersion.
We note that a
complete understanding of the ARPES experiments
may require a multi-band model, as well as inclusion of
dielectric and charging effects.

To carry out an Ising type expansion\cite{gel00,rajiv_sq} for this system
at $T=0$ we consider the Hubbard-Ising model
with the following Hamiltonian:
%Some details on the calculations and graphs involved.
%We have also carried out an Ising type expansion using a
%linked-cluster method.\cite{gel00}
%Similar expansions were previously done
%for the Hubbard model on the square lattice.\cite{rajiv_sq}
%To perform the series expansion, one needs to introduce an Ising
%interaction into the Hubbard Hamiltonian,
%and divide the Hamiltonian into an unperturbed
% Hamiltonian ($H_0$)
%and a perturbation ($H_1$) as follows,
%
%%%%%%%%%%%%%%  FIGURE  %%%%%%%%%%%%%%%%%%%%%%%%%%%%%%%%%%%%%%%%%%%%%%
\begin{figure}[!htb]
\begin{center}
  \includegraphics[width=7.5cm,bbllx=17,bblly=183,bburx=561,bbury=641,angle=0]{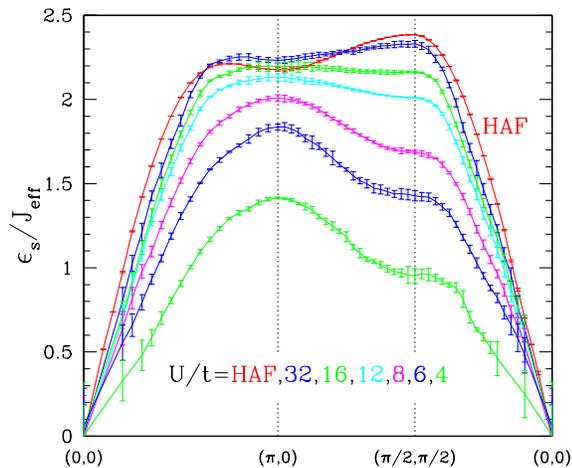}
  \caption{\label{fig1} (Color online) Magnon dispersion curve  along selected directions in the Brillouin zone
   for various values of $U/t$, expressed in units of $J_{\rm eff}=4t^2/U$.
Results for the Heisenberg antiferromagnet are shown as a red curve.}
\end{center}
\end{figure}
%%%%%%%%%%%%%%%%%%%%%%%%%%%%%%%%%%%%%%%%%%%%%%%%%%%%%%%%%%%%%%%%%%%%%%%%
%%%%%%%%%%%%%%  FIGURE  %%%%%%%%%%%%%%%%%%%%%%%%%%%%%%%%%%%%%%%%%%%%%%
\begin{figure}[!htb]
\begin{center}
  \includegraphics[width=7.5cm,bbllx=20,bblly=160,bburx=550,bbury=640,angle=0]{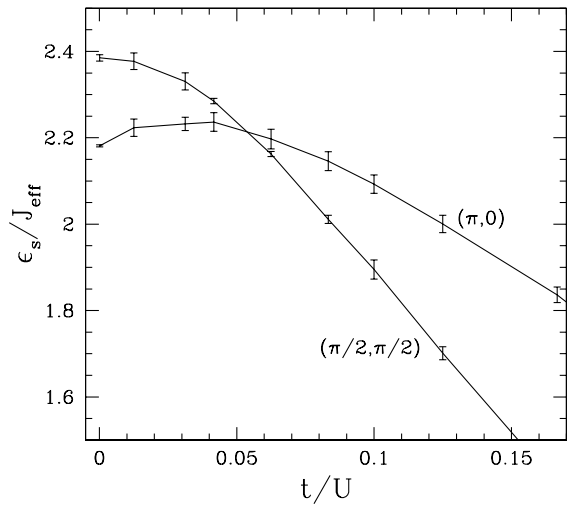}
  \caption{\label{fig2} Magnon energies at two wavevectors along
the antiferromagnetic zone boundary as a function of $t/U$.}
\end{center}
\end{figure}
%%%%%%%%%%%%%%%%%%%%%%%%%%%%%%%%%%%%%%%%%%%%%%%%%%%%%%%%%%%%%%%%%%%%%%%%
%
\bea
H &=& H_0 + \lambda H_1  \nonumber \\
H_0 &=& J/4 \sum_{\langle {\bf ij}\rangle}  ( \sigma_{\bf i}^z \sigma_{\bf j}^z  + 1)
+  \sum_{{\bf i}} [ U (n_{{\bf i}\uparrow} - \half ) (n_{{\bf i}\downarrow} - \half )
\nonumber \\
&& + h (-1)^{\bf i} \sigma_{\bf i}^z ]
\nonumber  \\
H_1 &=& - \sum_{\langle {\bf ij}\rangle} [  J ( \sigma_{\bf i}^z \sigma_{\bf j}^z  + 1)/4
    + t  ( c_{{\bf i}\sigma}^\dag c_{{\bf j}\sigma} + {\rm h.c.} ) ] \nonumber \\
&&  -h \sum_{{\bf i}} (-1)^{\bf i} \sigma_{\bf i}^z
\nonumber
\eea
where $\sigma_{\bf i}^z = n_{{\bf i}\uparrow} - n_{{\bf i}\downarrow}$,
and $\lambda$ is the expansion parameter. The Ising interaction $J$ is,
in principle, an adjustable parameter but here is chosen to
 be $4t^2/U$.
The strength of the staggered field $h$ can be varied to improve convergence.
Note that the full Hubbard model is recovered
at $\lambda=1$, at which point
the extra terms cancel between $H_0$ and $H_1$. On the other hand for
$\lambda<1$, there is an Ising-like anisotropy in the system, which favors
a Neel state and induces a gap in the spectrum.
The limit $\lambda=0$ corresponds to the Ising model,
with the usual N\'eel states being the two unperturbed ground states.

We have extended the linked cluster method \cite{gel00} to the
spectra of the Hubbard-Ising
models. At $\lambda=0$, the model has a very simple excitation spectrum.
Above the two ground states, all single spin-flip states
(or all states with a single hole for the hole spectra)
are degenerate with each other. We construct an orthogonality
transformation, which order by order in powers of $\lambda$, decouples
these single spin-flip states from the rest of the Hilbert space.
This procedure amounts to finding the combination of one-flip states
with others that remains an eigenstate as $\lambda$ changes from zero.
For a translationally invariant system the resulting block diagonal
Hamiltonian in the one particle subspace is diagonalized by
Fourier transformation.

To set the normalizations for our calculations, we begin with the
definition for the dynamic structure factor
\bea
S^{\alpha\beta}({\bf q},\omega)&=&\frac{1}{2\pi N} \sum_{i,j}\int_{-\infty}^{\infty}
\exp[i(\omega t + {\bf q}\cdot ({\bf r}_i - {\bf r}_j))] \nonumber \\
&&\langle S^{\alpha}_j (t)S^{\beta}_i (0)\rangle dt
\label{eq2}
\eea
We define the single magnon contribution to the dynamical transverse
structure factor as
\begin{equation}
S^{XX} ({\bf q},\omega) + S^{YY} ({\bf q},\omega) =A_s\delta(\omega-\epsilon_s({\bf q}))+B_s({\bf q},\omega)
\end{equation}
Here, $\epsilon_s({\bf q})$ gives the dispersion for the magnons and $A_s({\bf q})$
defines the weights for the magnons. The quantity $B_s({\bf q},\omega)$ defines
the multiparticle contributions.
Series expansions have been calculated for the spectra $\epsilon_s({\bf q})$ and
the spectral weights $A_s({\bf q})$ up to order $\lambda^{11}$.

Similarly, for the hole-excitations, we define the spectral function
\bea
A({\bf q},\omega) &=& %\sum_{\sigma=\uparrow,\downarrow}
\frac{1}{2\pi N} \sum_{i,j}\int_{-\infty}^{\infty}
\exp[i(\omega t + {\bf q}\cdot ({\bf r}_i - {\bf r}_j))] \nonumber \\
&&  \langle c^{\dagger}_{j,\sigma} (t) c_{i,\sigma} (0)\rangle dt
%&& \langle c^{\dagger}_{j,\sigma} (t) c_{i,\sigma} (0)\rangle dt
\label{eq3}
\eea
We define the single hole contribution to the spectral function as
\begin{equation}
A({\bf q},\omega) =A_h\delta(\omega-\epsilon_h({\bf q}))+B_h({\bf q},\omega)
\end{equation}
Series expansions are calculated for the hole spectra $\epsilon_h({\bf q})$ and
the spectral weights $A_h({\bf q})$ up to order $\lambda^{11}$.

The gap in the spectrum closes at $\lambda=1$, when spin rotational symmetry
is restored in the model. This causes power-law singularities in certain
properties of the model \cite{huse}.
Hence, the $\lambda=1$ limit needs to be dealt
with by series extrapolation methods. We use the method of integrated
differential approximants,  well known from the study of
critical phenomena \cite{gut}, to calculate various properties at $\lambda=1$.

%%%%%%%%%%%%%%  FIGURE  %%%%%%%%%%%%%%%%%%%%%%%%%%%%%%%%%%%%%%%%%%%%%%
\begin{figure}[!htb]
\begin{center}
  \includegraphics[width=7.5cm,bbllx=20,bblly=185,bburx=565,bbury=640,angle=0]{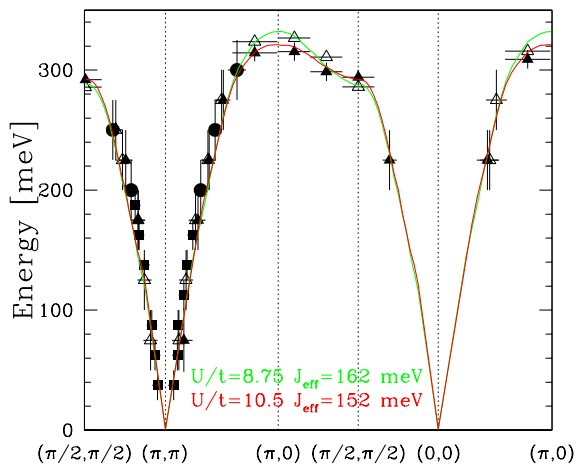}
  \caption{\label{fig_LaCuO} (Color online)
Fits to the spectra of the material  La$_2$CuO$_4$:  experimental
spectra at 10K (open symbols)   fit by $U/t=8.75\pm .7$, $J_{\rm eff}=162\pm 3 {\rm meV}$ (green curve);
 experimental spectra at 295K (solid symbols) fit with $U/t=10.5\pm .7$,
$J_{\rm eff}=152\pm 3 {\rm meV}$ (red curve).
}
\end{center}
\end{figure}
%%%%%%%%%%%%%%%%%%%%%%%%%%%%%%%%%%%%%%%%%%%%%%%%%%%%%%%%%%%%%%%%%%%%%%%%
%%%%%%%%%%%%%%  FIGURE  %%%%%%%%%%%%%%%%%%%%%%%%%%%%%%%%%%%%%%%%%%%%%%
\begin{figure}[!htb]
\begin{center}
  \includegraphics[width=7.5cm,bbllx=20,bblly=185,bburx=565,bbury=640,angle=0]{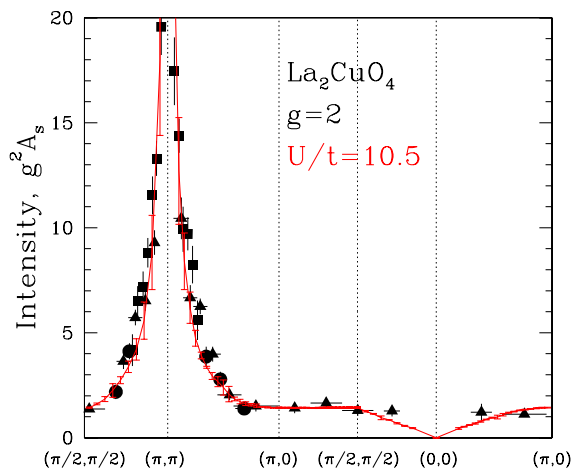}
  \caption{\label{fig_LaCuO_wt} (Color online)
%Comparison of magnon spectral weight $Z_d \pi g^2 A_s/2$
Comparison of magnon spectral weight $g^2 A_s$
for $U/t=10.5$ with neutron scattering intensity\cite{Radu}
for  La$_2$CuO$_4$ at 295K.
}
\end{center}
\end{figure}
%%%%%%%%%%%%%%%%%%%%%%%%%%%%%%%%%%%%%%%%%%%%%%%%%%%%%%%%%%%%%%%%%%%%%%%%

In Figure \ref{fig1}, we show the calculated magnon dispersion,
in units of $J_{\rm eff}=4t^2/U$, along
selected directions in the Brillouin zone, for several values
of $t/U$. The results\cite{HAF} for the Heisenberg model are also shown.
While the long wavelength spin-wave velocity is gradually
reduced with increasing $t/U$, other dramatic changes
arise along the antiferromagnetic zone boundary. The magnon
energy at wavevector ($\pi,0$) at first rises briefly before
it begins to decrease with increasing $t/U$. On the other hand,
the magnon energy at ($\pi/2,\pi/2$) decreases sharply with
increasing $t/U$. Both of these are plotted in Figure \ref{fig2}, where one
can see that they cross at a relatively small value of $t/U= 0.053(5)$.

In Figures \ref{fig_LaCuO} and \ref{fig_LaCuO_wt} we show fits  to
the magnon spectra and spectral weight of La$_2$CuO$_4$.
Since we have not done any finite temperature calculations,
we try to fit the spectra at different temperatures by
effective $U$ and $t$ values.
We find that
the spectrum at 10K is fitted well by $U/t=8.75\pm 0.7$ and
$J_{\rm eff}=162\pm 3$ meV, while that at 295K is fitted best by
$U/t=10.5\pm 0.7$ and $J_{\rm eff}=152\pm 3$ meV.
These results suggest that as the temperature is increased the
effective exchange constant decreases whereas the effective $U/t$
ratio increases.
Assuming that the 10K data are essentially at $T=0$,
we obtain bare parameters of $U=3.1{\rm eV}$ and $t=0.35{\rm eV}$.

The spectral weights are only measured at 295K,
hence we show a fit to the calculated spectral weights at the larger
$U/t$ ratio. In any case, the spectral weights are not very
sensitive to the $U/t$ ratio. The relative fit is excellent. To
get a measure of the absolute fit, we first note that the integrated
one-magnon spectral weight over the entire zone was \cite{Radu_weight}
found from experiment to be $0.36\pm 0.09$,
in the normalization where total transverse spectral weight is $0.5$.
With this normalization, our more accurate calculations for the Heisenberg model
give an integrated transverse one-magnon spectral weight of $0.419(2)$.
The one-magnon spectral weights decrease slightly with decreasing $U/t$, and
 are also much less accurate, giving $0.40(8)$ for $U/t=10$. Hence, the
results agree with experiments well within the uncertainties.
We should note, however, that this agreement is very much dominated by the
$1/q$ dependent behavior near the antiferromagnetic wavevector (where $q$
is the deviation from the antiferromagnetic wavevector), which does not depend much
on the microscopic model. The best place to look for
multimagnon excitations and a more sensitive comparison with microscopic
models is the spectral weight along the antiferromagnetic zone boundary.
For the Heisenberg model, the multiparticle spctral weight is largest at
($\pi,0$), where it is about $40$ percent of the total transverse spectral weight.
We hope our work may motivate more accurate measurements of multiparticle
spectral weights along the antiferromagnetic zone boundary.
%If the experimental spectra are normalized\cite{Radu_weight}
%as $Z_d\pi g^2 A_s/2$, and we set the g-factor $g=2$, the fitting requires
%a  multiplicative
%renormalization factor $Z_d=0.65$.
% However, there is an overall renormalizaion factor of $0.65$ used to obtain the fit.
%This could be partially related to a reduction in spectral weight
%with temperature and partially to uncertainties in measurement of
%absolute intensities \cite{Radu}.

%%%%%%%%%%%%%%  FIGURE  %%%%%%%%%%%%%%%%%%%%%%%%%%%%%%%%%%%%%%%%%%%%%%
\begin{figure}[!htb]
\begin{center}
  \includegraphics[width=7.5cm,bbllx=20,bblly=180,bburx=565,bbury=660,angle=0]{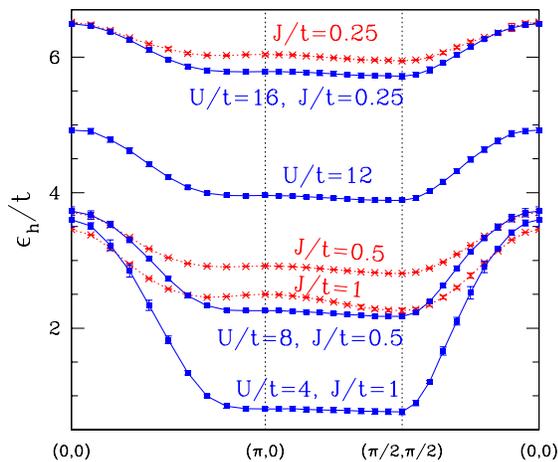}
  \caption{\label{fig_1h_dispersion} (Color online) Single hole dispersion
  in the square-lattice
half-filled Hubbard model along selected directions in momentum
space for several values of $t/U$ (squares). Also shown as
crosses is the dispersion for
the $t$-$J$ model with comparable $t/J_{\rm eff}$ ratios.}
\end{center}
\end{figure}
%%%%%%%%%%%%%%%%%%%%%%%%%%%%%%%%%%%%%%%%%%%%%%%%%%%%%%%%%%%%%%%%%%%%%%%%

%%%%%%%%%%%%%%  FIGURE  %%%%%%%%%%%%%%%%%%%%%%%%%%%%%%%%%%%%%%%%%%%%%%
%\begin{figure}[!htb]
%\begin{center}
%  \includegraphics[width=7.5cm,bbllx=20,bblly=180,bburx=565,bbury=660,angle=0]{1h_S1p.ps}
%  \caption{\label{fig_1h_weight} (Color online) Single hole spectral weight $A_h$
%  along selected directions in momentum
%space for several values of $t/U$.}
%\end{center}
%\end{figure}
%%%%%%%%%%%%%%%%%%%%%%%%%%%%%%%%%%%%%%%%%%%%%%%%%%%%%%%%%%%%%%%%%%%%%%%%

The data on two other systems of square-lattice antiferromagnets
(CuDCOO)$_2\cdot 4$D$_2$O (CFTD) \cite{Ronnow}
and Cu(II) spins of
Sr$_2$Cu$_3$O$_4$Cl$_2$ \cite{Kim} have much larger $U/t$ ratios
and are well fitted by the Heisenberg model.
Allowing $U$ to vary, (CuDCOO)$_2\cdot 4$D$_2$O (CFTD) can
be well fitted by the Hubbard model with $U/t=50$ and $U=3.9$eV.
On the other hand, the experimental data on Sr$_2$Cu$_3$O$_4$Cl$_2$ have substantial uncertainties and anisotropies,
so one can not get a reliable estimate for $U/t$. If one assumes the $U$ value for
it is about $4{\rm eV}$, one estimates $U/t\approx 40$.
% , can not be fit well with the Hubbard model with a reasonable $U$ value
% and suggests that the effective second neighbor interactions
% between the Cu(II) spins may be
% more significant in this material. This maybe becaue the Cu(II) spins
% couple indirectly with each other through the more strongly coupled
% Cu(I) spins and, we suggest that,
% this leads to more substantial second neighbor interactions.

In Figure \ref{fig_1h_dispersion},  we show the single hole excitation spectra
along selected directions in momentum space at different values
of $t/U$. Plots are also shown for the $t$-$J$ model\cite{tJ}, with comparable
$t/J$ values. It is
evident that the band-width scales with $J$ but is much larger
than in the t-J model.
The reason for this is the effective same-sublattice hopping generated
in the Hubbard model in the order $t^2/U$ which is not included in the $t$-$J$
model.
In all cases, the minimum of
the hole energy remains at ($\pi/2,\pi/2$) and the dispersion
along the line ($\pi/2,\pi/2$) to ($\pi,0$) remains weak. As mentioned above,
this is
in contrast to the observed single hole dispersion\cite{ARPES} in the material
Sr$_2$CuO$_2$Cl$_2$.
Furthermore, our calculated spectral weights at ($\pi/2,\pi/2$)
and ($\pi,0$) are very similar.
It has been argued that even small same-sublattice hopping terms
can significantly change the shape of the dispersion curves
and bring them closer to those observed in ARPES measurements \cite{ARPES},
% but it will be difficult to
they may also help
reconcile the measured spectral weights in the undoped cuprates \cite{oleg}.
%where ($\pi,0$) and ($\pi/2,\pi/2$) appear
%radically different, with the Hubbard model where they are so similar.

%\section{Conclusions}

In conclusion, we find that allowing for charge fluctuations by
making $t/U$ finite allows us to understand the antiferromagnetic
zone-boundary excitations in the material La$_2$CuO$_4$ very well.
% However, it does not allow us to understand the single hole spectra in
% the undoped cuprate materials.
However, it appears that the simple one-band Hubbard model considered here
is unable to explain the single hole spectra in the undoped cuprate materials.
%Among other things, the inclusion of multiple bands and charging/dielectric
%effects may be needed.

%Acknowledgements

We would like to thank R. Coldea for useful discussion and sending us the experimental data.
This work is supported by the
Australian Research Council and the US National Science
Foundation grant number DMR-0240918 (RRPS).
 We are grateful for the computing resources provided
 by the Australian Partnership for Advanced Computing (APAC)
National Facility and by the
Australian Centre for Advanced Computing and Communications (AC3).


% \newpage

\end{document}